\documentclass[a4paper]{jpconf}
\usepackage{iopams}
\usepackage{amsmath, amssymb}
\usepackage{slashed}
\usepackage{nicefrac}
\usepackage{graphicx}
\usepackage{subfigure}
\usepackage{verbatim}
\usepackage{mathrsfs}

\usepackage{times}
\usepackage{mathptmx}

\usepackage{cite}
\usepackage[colorlinks, citecolor=blue, linkcolor=blue, urlcolor=blue]{hyperref}

\newcommand{\figlab}[1]{\label{fig:#1}}
\newcommand{\tablab}[1]{\label{tab:#1}}
\newcommand{\equlab}[1]{\label{eq:#1}}
\newcommand{\seclab}[1]{\label{sec:#1}}

\newcommand{\figref}[1]{Fig.\,\ref{fig:#1}}
\newcommand{\tabref}[1]{Tab.\,\ref{tab:#1}}
\newcommand{\equref}[1]{Eq.\,\eqref{eq:#1}}
\newcommand{\secref}[1]{Sec.\,\ref{sec:#1}}

\newcommand{\subfigref}[2]{\figref{#1}\,(#2)}
\newcommand{\subfigsref}[2]{\figsref{#1}\,(#2)}
\newcommand{\figsref}[1]{Figs.\,\ref{fig:#1}}

\renewcommand{\vec}{\mathbf}

\newcommand{\im}{\mathrm{i}}
\newcommand{\diff}{\mathrm{d}}
\newcommand{\one}{1}

\newcommand{\smat}{\mathcal{S}}
\renewcommand{\smat}{\mathscr{S}}

\newcommand{\sci}[2]{{#1}{\times}{10^{#2}}}
\newcommand{\unit}[1]{~\text{#1}}
\newcommand{\abs}[1]{\left|\,{#1}\,\right|}

\newcommand{\scalar}[3][\mu]{#2_#1 #3^#1}

\newcommand{\avg}{\overline}
\newcommand{\compl}{\bar}

\begin{document}

%\title{Bethe--Heitler Pair Creation in a Bichromatic or an Elliptically Polarized Laser Field}
\title{Nonlinear Bethe--Heitler Pair Creation in an Intense Two-Mode Laser Field}

\author{Sven Augustin$^{1,2}$ and Carsten M\"uller$^{1,2}$}

\address{$^1$ Institut f\"ur Theoretische Physik I, Heinrich-Heine-Universit\"at, 40225 D\"usseldorf, Germany}
\address{$^2$ Max-Planck-Institut f\"ur Kernphysik, Saupfercheckweg 1, 69117 Heidelberg, Germany}

\ead{sven.augustin@mpi-hd.mpg.de}

\begin{abstract}
We investigate electron-positron pair creation in the interaction of a nuclear Coulomb field and a highly intense two-mode laser field.
For bichromatic laser fields, we examine the differences arising for commensurable and incommensurable frequencies in a continuous variation of the laser frequency ratio and the quantum interference effects, which may occur in the commensurable case.
We show that the interference manifests in the angular distributions and the total pair-production rates of the created particles.
Additionally, by varying the amplitudes of the two modes we study pair creation in a monochromatic laser wave of arbitrarily elliptical polarization.
%
%Additionally, a transition from a single purely linearly polarized laser wave to a circularly one is shown, covering the elliptical polarizations in between.
\end{abstract}

%\twocolumn

%\newpage
\section{Introduction}
%\nocite{augustin}

The creation of matter from laser light has already been theoretically investigated \cite{reiss, nikishov, narozhny-mono, yakovlev} shortly after the realization of the first laser itself \cite{maiman}. Recently this interest has experienced a revival \cite{ehlotzky-review, dipiazza-review}, due to the large and still ongoing progress in high-intensity laser technology.
Additionally, the experimental feasibility of electron-positron ($e^-e^+$) pair production by multiphoton absorption was demonstrated by a pioneering experiment at the Stanford Linear Accelerator Center (SLAC) \cite{e144-1997, e144-1999}, applying the nonlinear Breit--Wheeler process \cite{reiss, nikishov, narozhny-mono}.
In a similar manner, the nonlinear Bethe--Heitler process
\begin{equation}
Z + n \omega \to Z + e^- + e^+
\end{equation}
is in principle accessible by modern experimental techniques, e.g., by using the highly relativistic nuclear beam from the Large Hadron Collider (LHC) at CERN in conjunction with a counterpropagating highly intense laser beam. In the nuclear rest frame, the laser frequency and intensity are largely amplified by a relativistic Doppler shift, reaching the levels required for pair production.

The prospect of an experimental test has led to further theoretical investigations dedicated to the nonlinear Bethe--Heitler effect.
On the one hand the various field parameter regimes have been studied by calculating total and differential pair-production rates (e.g., \cite{cm-circ, avetissian, cm-lin, sieczka, kuchiev, fillion}), while on the other hand more specialized features were examined, such as the effects of the electron spin \cite{tomueller} and the nuclear recoil \cite{smueller, krajewska-recoil}.
In all these studies the laser field was assumed to be a monochromatic plane wave with either linear or circular polarization.

The subject of this contribution is Bethe--Heitler pair creation in a two-mode laser field. In particular, the modes may oscillate with different frequencies, leading to a bichromatic laser field:
\begin{equation}
Z + n_1 \omega_1 + n_2 \omega_2 \to Z + e^- + e^+.
\end{equation}
While in general arbitrary combinations of the laser frequencies $\omega_i$ are allowed, so far mainly two special cases have been investigated:
largely differing frequencies with $\nicefrac{\omega_1}{\omega_2} \gtrsim 10^2$ \cite{loetstedt, dipiazza}
and, as only there quantum path interference may occur,
commensurable frequencies, i.e., frequencies of rational ratio \cite{roshchupkin, krajewska-phase, krajewska-symm, augustin}.
In \cite{krajewska-phase, krajewska-symm, augustin} it could be concluded that the relative phase between the two laser modes distinctly influences the angular spectrum of the created pairs. In \cite{krajewska-phase, krajewska-symm} both field modes were assumed to be linearly polarized along the same direction.
%
%In \cite{krajewska-phase, krajewska-symm} a laser field geometry where both modes were linearly polarized along the same direction. %and of equal intensity was used.
%
For other examples of interference effects in field-induced pair creation, we refer the interested reader to \cite{narozhny, fedorov, voitkiv, cheng-grobe, jiang-grobe, dumlu-Dunne, akkermans-Dunne}.

In the present contribution, we will extend our previous study \cite{augustin} to incommensurable frequencies, again assuming both field modes to be linearly polarized with mutually orthogonal polarization vectors and propagating in the same direction.
Additionally, we will examine further aspects of the interference in the commensurable case by comparing a laser pair combination from our previous study to a new one with a higher number of photons, emphasizing on the features arising from this increase.
Finally, we investigate the special case of a monochromatic laser wave, by a variation of the individual intensities of the two laser modes. This allows a transition from a single linearly polarized to a circularly polarized laser wave, covering elliptical polarizations in between.

%We will show that the interference modifies the angular distribution of the created particles in form of peak positions shifts. It may also lead to an increase (or a decrease) of the total pair-production rate. Both modifications depend on the relative phase between the two laser modes. The phase dependence of the total pair production rate can be intuitively explained by the electric field of the combined laser wave.

%In order to fully account for the interaction of the leptons with the laser field we apply the Volkov solutions to the Dirac equation as basis states of an $\smat$-matrix formalism in the Furry picture, wherein the treatment of the nuclear Coulomb field is restricted to the lowest order of perturbation theory.
%
%The results are discussed in the nuclear rest frame.

%Besides the aforementioned studies, several related works are worth mentioning:
%
%Interference effects have been studied in $e^-e^+$ pair creation by a highly energetic non-laser photon in the presence of a bichromatic laser field of commensurable frequencies \cite{}, as well as other types of interference effects in field-induced pair production \cite{narozhny, dumlu-Dunne, akkermans-Dunne, cheng-grobe, jiang-grobe}.

The present article is organized as follows. First, we will outline our calculational approach in \secref{calc}.
Into the $\smat$-matrix describing the nonlinear Bethe--Heitler process we insert the Volkov solutions of the Dirac equation for a bichromatic laser field with linearly polarized modes of orthogonal field vectors. From this, an expression for the total pair-production rate is derived, containing a six-fold integral over the momenta of the created particles and a four-fold sum over photon numbers, which both can effectively be reduced by one due to constraints from energy conservation. 
The remaining integrations are performed numerically, leading to the results presented in \secref{results}, where we show total pair-production rates or pair-production rates differential in the polar emission angle.
%
%The different combination of indices in the aforementioned four-fold sum, with the categories introduced in \secref{notation}, will be studied by means of their respective partial contributions to the summed-up pair-production rates.
%
Section\,\ref{sec:commen} is devoted to the case of commensurable laser frequencies for which quantum interference effects may occur.
%
%Besides the comparison of two different photon orders for the same total photon energy, we will show the influence of the relative phase between the laser modes and give an intuitive explanation for the latter using the electric field of the combined laser wave.
%
We will compare rates obtained for two different photon orders but identical total photon energy and discuss the influence of the relative phase between the laser modes. 
%
%The difference between commensurable and incommensurable frequencies will be shown in a variation of the frequency ratio of the two laser modes.
%
In \secref{incomm} the difference between commensurable and incommensurable field frequencies for the nonlinear Bethe--Heitler process will be analyzed by showing pair-creation rates in a variation of the frequency ratio of the two laser modes.
%
%The special case of a monochromatic laser field with elliptical polarization will be discussed in form of two different variations of the ellipticity parameter from circular to linear polarization.
%
The special case of pair creation in a monochromatic laser field with elliptical polarization will be discussed in \secref{ellip}. The dependence of the total pair-creation rate on the ellipticity will be examined by varying the field polarization continuously from circular to linear in two different ways.
The conclusions that can be drawn from our study are summarized in \secref{concl}.

%\newpage
\section{Theoretical Framework} \seclab{calc}
%\subsection{Volkov Solutions for a Bichromatic Field}
%\subsection{Transition Amplitude and Pair-Production Rates}
\subsection{Pair-Production Amplitude and Rate for a Two-Mode Field}

We model pair creation in the superposition of a nuclear field and a laser field as a transition from a negative continuum state $\Psi^{(+)}$ to one of the positive continuum $\Psi^{(-)}$, induced by the nuclear Coulomb potential
\begin{equation}
A_\text{N} = \frac{Ze}{\abs{\vec{r}}}.
\end{equation}
Treating the nuclear Coulomb field in the lowest order of perturbation theory leads to the pair-creation amplitude \cite{yakovlev, cm-circ, cm-lin, sieczka, kuchiev}
\begin{equation} \equlab{smat}
\smat = \frac{\im \, Z e^2}{\hbar c} \int \compl\Psi^{(-)} \gamma_0 \Psi^{(+)} \, \frac{\diff^4 \! x}{\abs{\vec{r}}},
\end{equation}
in the rest frame of the nucleus.
As continuum wave functions, in order to fully account for the interaction of the leptons with the laser field, we use the Volkov solutions \cite{volkov} for electrons and positrons, labelled by the sign of their charge $-$ and $+$, respectively,
\begin{equation} \equlab{volkov}
\Psi_{p_\pm,s_\pm}^{(\pm)} = N_\pm
\left( \one \pm \frac{e \slashed{\kappa} \slashed{A}}{2c \scalar{\kappa}{p_\pm}} \right)
\exp\!{\left( \frac{\im}{\hbar} S^{(\pm)} \right)} ~u_{p_\pm,s_\pm}^{(\pm)},
\end{equation}
containing the normalizer $N$, the action $S$,
%%
%\begin{equation} \equlab{action}
%S^{(\pm)} = \pm \scalar{p}{x} + \frac{e}{c\scalar{p}{k}} \int^\eta \! \left[ \scalar{p}{A}(\tilde\eta) \mp \frac{e}{2c} A^2(\tilde\eta) \right] \diff\tilde\eta,
%\end{equation}
%%
and the free Dirac spinor $u$ with the respective particle's momentum $p$ and spin $s$.
Herein the positive elementary charge $e$
%the four-gradient $\partial = \left( \frac{1}{c}\frac{\partial}{\partial t}, -\nabla \right)$, 
%and the wave vector $k = (\nicefrac{\omega}{c}, \vec{k})$ are used, 
and the unit vector in wave propagation direction $\kappa$ are used,
%and the four-dimensional space-time coordinate $x = (ct, \vec{r})$ are used, 
and Feynman slash notation $\slashed{A} = \scalar{\gamma}{A}$ is applied.
%
%According to the particle charge, we distinguish between electron and positron by the subscripts $-$ and $+$, respectively, for the momentum, the spin, and the normalizer.
%
The Volkov wave functions are gained as exact solutions of the Dirac equation %\cite{dirac}
%%
%\begin{equation} \equlab{dirac}
%\left( \im \hbar \slashed{\partial} + \frac{e}{c} \slashed{A} - mc \right) \Psi = 0
%\end{equation}
%%
for an electron moving in the field of an electromagnetic plane wave in vacuum,
where the vector potential $A$, given in Lorenz gauge $\scalar{\partial}{A} = \scalar{\kappa}{A} = 0$, depends only on a phase variable $\eta$. % = \scalar{k}{x} = \omega t - \vecprod{\vec{k}}{\vec{r}}$.
%%
%\begin{equation}
%\eta = \scalar{k}{x} = \omega t - \vecprod{\vec{k}}{\vec{r}}.
%\end{equation}
%%

%We model pair production as a transition from a negative continuum Volkov state $\Psi^{(+)}$ to one of the positive continuum $\Psi^{(-)}$, induced by the nuclear Coulomb potential
%%
%\begin{equation}
%A_\text{N} = \frac{Ze}{\abs{\vec{r}}}.
%\end{equation}
%%
The Volkov wave functions can be inserted into the pair-creation amplitude from \equref{smat}, leading to
%%
%\begin{equation}
%\smat = \frac{\im \, Z e^2}{\hbar c} \int \compl\Psi_{p_-,s_-}^{(-)} \gamma_0 \Psi_{p_+,s_+}^{(+)} \, \frac{\diff^4 \! x}{\abs{\vec{r}}}.
%\end{equation}
%%
%we obtain
%
\begin{equation} \equlab{smat2}
\begin{split}
\smat
&=
N_- N_+ 
\frac{\im Z e^2}{\hbar c} 
\int \! \frac{\diff^4 \! x}{\abs{\vec{r}}} \,
G
\exp\!{\left(\frac{\im}{\hbar} \left(-S^{(-)} + S^{(+)} \right)\right)},
\\
G
&= 
\compl{u}_{p_-,s_-}^{(-)}
\left(
\one - \frac{e \slashed{A} \slashed{\kappa}}{2c \scalar{\kappa}{p_-}}
\right)
\gamma_0
\left(
\one + \frac{e \slashed{\kappa} \slashed{A}}{2c \scalar{\kappa}{p_{\vphantom{-}\smash{+}}}} \right)
u_{p_+ ,s_+}^{(+)},
\end{split}
\end{equation}
where the abbreviation $G$ contains all $\gamma$-matrices.

%In the present study
The laser field $A$ %$A = A_1 + A_2$ 
is defined as superposition of two plane waves, %$A_i = a_i \cos(\eta_i +  \varphi_i)$ for $i=1 ~\text{or}~ 2$,
\begin{equation} \equlab{laserwave}
A = A_1 + A_2
\quad
\text{with}
\quad
A_i = a_i \cos(\eta_i + \varphi_i)
\quad
(i=1 ~\text{or}~ 2),
\end{equation}
with relative phases $\varphi_i$,
phase coordinates $\eta_i = (\nicefrac{\omega_i}{c}) \, \scalar{\kappa}{x}$, where the direction of propagation $\kappa = \left( 1, 0, 0, 1 \right)$ is shared among the wave vectors $k_i = (\nicefrac{\omega_i}{c}) \kappa$,
and perpendicular field vectors $a_i$, given by $a_1 = \left( 0, 1, 0, 0 \right) \abs{\vec{a}_1\!}$ and $a_2 = \left( 0, 0, 1, 0 \right) \abs{\vec{a}_2\!}$,
%
%\begin{equation}
%a_1 = \left( 0, 1, 0, 0 \right) \abs{\vec{a}_1\!}
%\quad\text{and}\quad
%a_2 = \left( 0, 0, 1, 0 \right) \abs{\vec{a}_2\!},
%\end{equation}
%
measuring their absolute amplitudes using the dimensionless intensity parameters
\begin{equation} \equlab{xi}
\xi_i = \frac{e}{mc^2} \frac{\abs{\vec{a}_i\!}}{\sqrt{2}},
\end{equation}
with the electronic rest mass $m$ and the speed of light in vacuum $c$.

Due to the chosen field geometry, functions of the two phase coordinates are still separable when the laser amplitude is squared.
%
%Using the abbreviation $\eta'_i = \eta_i  + \varphi_i$, 
We can thus give the action as sum over the laser modes:%, with the necessary integrations over the individual $\diff\tilde\eta_i$ already performed:
\begin{equation} \equlab{actionsum}
S^{(\pm)}
= \pm \scalar{p}{x} + \sum_{i=1}^{2}
\frac{e}{c\scalar{p}{k_i}}
\left[
%\scalar{p}{a_i} \sin(\eta'_i)
\scalar{p}{a_i} \sin(\eta_i  + \varphi_i)
\mp \frac{e}{4c} a_i^2 \left(
%\frac{\sin(2\eta'_i)}{2}
\frac{\sin(2[\eta_i  + \varphi_i])}{2}
%+ \eta'_i
+ (\eta_i  + \varphi_i)
\right)
\right].
\end{equation}
%

%\subsection{Transition Amplitude and Pair-Production Rates}

Upon insertion of the action from \equref{actionsum} and the laser fields from \equref{laserwave} into the pair-creation amplitude from \equref{smat2} we obtain for each laser mode a set of three functions periodic in the respective phase coordinate $\eta_i$. These functions can be expanded in Fourier series and the resulting coefficients are built from generalized Bessel functions \cite{reiss-genbes}.
With this expansion we can write the amplitude as a summation over two indices, which can be interpreted as counts for the number of photons taken from each of the two modes:
\begin{equation} \equlab{smatrix}
\smat = 
\frac{\im Z e^2 m c}{\hbar c \sqrt{q_-^0 q_+^0}} 
\sum_{n_1, n_2}
M_{p_- p_+}^{(n_1, n_2)}
\int \! \frac{\diff^4 \! x}{\abs{\vec{r}}} \,
\exp\!{\left(\frac{\im}{\hbar} \scalar{x}{Q_{(n_1, n_2)}}\right)}.
\end{equation}
Here we have introduced
the matrix element $M_{p_- p_+}^{(n_1, n_2)}$, which consists of all slashed quantities and the aforementioned Fourier coefficients,
the normalizers $N_\pm = \sqrt{\nicefrac{m c}{q_\pm^0}}$,
%%
%\begin{equation} 
%N_\pm = \sqrt{\frac{m c}{q_\pm^0}},
%\end{equation}
%%
and the momentum transfer to the nucleus $Q_{(n_1, n_2)} = q_+ + q_- - n_1 \hbar k_1 - n_2 \hbar k_2$,
%%
%\begin{equation}
%Q_{(n_1, n_2)} = q_+ + q_- - n_1 \hbar k_1 - n_2 \hbar k_2,
%\end{equation}
%%
where the latter two are defined using the effective momentum \cite{LL4}
\begin{equation}
q_\pm = p_\pm + \frac{e^2 \avg{\vec{A}^2}}{2c^2 \scalar{\kappa}{p_\pm}} \, \kappa,
\end{equation}
with the averaged squared laser amplitude $\avg{\vec{A}^2} 
= \frac{1}{2} \left(\abs{\vec{a}_1\!}^2 + \abs{\vec{a}_2}^2\right)
= \frac{m^2 c^4}{e^2} \left(\xi_1^2 + \xi_2^2\right).$
%%
%\begin{equation}
%\avg{\vec{A}^2} 
%= \frac{1}{2} \left(\abs{\vec{a}_1\!}^2 + \abs{\vec{a}_2}^2\right)
%= \frac{m^2 c^4}{e^2} \left(\xi_1^2 + \xi_2^2\right).
%\end{equation}
%%
%
The four-dimensional integral in \equref{smatrix} can be solved by using the Fourier transform of the Coulomb potential and a representation of the $\delta$-function for the integral in space and time, respectively \cite{bjorken-drell}.
%%
%\begin{align}
%&\int \! \diff^3 r \, \frac{1}{\abs{\vec{r}}} \, \e^{-\frac{\im}{\hbar} \vecprod{\vec{Q}\,}{\,\vec{r}}} = \frac{4 \pi \hbar^2}{\abs{\vec{Q}}^2}, \\
%&\int \! \diff x_0 \, \e^{\frac{\im}{\hbar} Q_0 \, x_0} = 2 \pi \hbar \, \delta(Q_0).
%\end{align}
%%
Note that by definition of $Q_{(n_1, n_2)}^0$ the newly introduced $\delta$-function ensures energy conservation.

Squaring the amplitude leads to a sum over four indices:
\begin{equation} \equlab{amplitude}
\abs{\smat}^2
=
\sum_{\substack{n'_1, n'_2\\n_1, n_2}}
\mathscr{P}_{[n_1,  n'_1, n_2,n'_2]},
\end{equation}
with the addends being the thereby defined partial contributions
\begin{equation} \equlab{partcontrib}
\mathscr{P}_{[n_1,  n'_1, n_2,n'_2]}
=
\frac{Z^2 e^4 m^2}{q_+^0 q_-^0}
~
32\pi^3 \hbar^3
~
\compl{M}_{p_- p_+}^{(n_1, n_2)} M_{p_- p_+}^{(n'_1, n'_2)}
\frac{cT}{Q_{(n_1, n_2)}^4} \, \delta\!\left(Q_{(n_1, n_2)}^0\right).
\end{equation}
Here we have used $Q_{(n_1, n_2)} = Q_{(n'_1, n'_2)}$ and $n_1 k_1 + n_2 k_2 = n'_1 k_1 + n'_2 k_2$
as enforced by the $\delta$-function,
and introduced the time $T$ from the squared $\delta$-function \cite{bjorken-drell}.
%
%\begin{equation} \equlab{deltasquared}
%\left[ 2\pi\hbar \, \delta\!\left(Q_{(n_1, n_2)}^0\right) \right]^2 
%= 2\pi\hbar \, \delta\!\left(Q_{(n_1, n_2)}^0\right) cT.
%\end{equation}
%
The product of the two matrix elements $\compl{M}_{p_- p_+}^{(n_1, n_2)}$ and $M_{p_- p_+}^{(n'_1, n'_2)}$ is a rather cumbersome summation of products of Dirac $\gamma$-matrices and thus shall not be shown here.

Finally, the partial contributions $\mathscr{P}$ enter the differential partial rates
\begin{equation} \equlab{partrate}
\diff^6 R_{[n_1,  n'_1, n_2,n'_2]} = 
\frac{1}{T} \sum_{s_+, s_-}
\mathscr{P}_{[n_1,  n'_1, n_2,n'_2]}
\frac{\diff^3 q_-}{(2\pi\hbar)^3} \frac{\diff^3 q_+}{(2\pi\hbar)^3},
\end{equation}
where we also summed over the final spin states and divided by the time $T$.
Using the $\delta$-function in \equref{partcontrib}, we can perform one integration analytically. The remaining integrals are calculated numerically to obtain angular differential and fully integrated partial rates. Additionally, the summation over photon numbers from \equref{amplitude} is performed to find differential and total rates.
Results from these computations are presented in \secref{results}.

%Even though partial rates are used throughout this work to interpret the obtained results, they are, in general, no experimental observables.
%
We emphasize that the partial rates $R_{[n_1,  n'_1, n_2,n'_2]}$ introduced above are, in general, no experimental observables.
%
%It is especially important to note that they may be negative and would then decrease the summed-up rate. 
%
It is especially important to note that, for commensurable laser frequencies, they may be negative and accordingly decrease the total rate
\begin{equation}
R = 
\sum_{\substack{n'_1, n'_2\\n_1, n_2}}
R_{[n_1,  n'_1, n_2,n'_2]}.
\end{equation}
The latter, on the other hand, is an experimentally measurable quantity and always positive. A negative contribution to the four-index sum may only arise for certain index combinations and will subsequently be interpreted as destructive interference between the two laser modes, just as a positive contribution from these particular index combinations will be understood as constructive interference.

%\newpage
\subsection{Terminology} \seclab{notation}

In the following, the terminology applied in the presentation of our results is briefly discussed.
The minimal number of photons from a single mode $i$ with frequency $\omega_i$ needed to create an electron-positron pair will be denoted by $\tilde{n}_i$. In the case of commensurable frequencies, a pair of two laser modes $(\tilde{n}_1, \tilde{n}_2)$ is then given by their respective minimal photon numbers if their total photon energies are identical, $E_\text{tot} = \tilde{n}_1 \hbar \omega_1 = \tilde{n}_2 \hbar \omega_2$, and thus it is indistinguishable whether $\tilde{n}_1$ photons were absorbed from the first mode or $\tilde{n}_2$ from the second mode.

\begin{table}[t]
\caption{\tablab{nota}
Types, conditions, and examples of the terms in the summation from \equref{amplitude}.}
\begin{center}
\begin{tabular}{@{}lcr@{}}
\br
Type of term & Condition & Example: $\left(4,8\right)$\\
\mr
Direct & $n_1 = n'_1, n_2 = n'_2 \enskip\text{and only one is not 0}$ & $\left[4,4,0,0\right]$ or $\left[0,0,8,8\right]$\\
Symmetrically mixed & $n_1 = n'_1 \neq 0 \enskip\text{and}\enskip n_2 = n'_2 \neq 0$ & $\left[2,2,4,4\right]$ \\
Interference (Asymmetrically mixed) & $n_1 \neq n'_1 \enskip\text{and}\enskip n_2 \neq n'_2$ & $\left[0,4,8,0\right]$ or $\left[0, 2, 8, 4\right]$\\
\br
\end{tabular}
\end{center}
\end{table}

It is useful to introduce three categories for the index combinations $\left[n_1, n'_1, n_2, n'_2\right]$ from \equref{amplitude}: \emph{direct} terms, \emph{symmetrically mixed} terms and \emph{asymmetrically mixed} terms. In \tabref{nota} the exact conditions and some examples are summarized.
The \emph{direct} terms stem solely from one of the two laser modes and are thus the only terms that could be measured individually in an experiment by turning off the respective other mode.
For the \emph{symmetrically mixed} terms the energy to overcome the pair-creation threshold is gained by taking a certain number of photons from the first mode and another number of photons from the second mode.
In contrast, for the \emph{asymmetrically mixed} terms this intuitive explanation in terms of photon numbers does not hold, instead they can be understood as stemming from interference of the two laser modes, which may only occur in the commensurable case discussed in \secref{commen}. As we will also see later, only these \emph{interference terms} are sensitive to a variation of the relative phases $\varphi_i$ from \equref{laserwave}.

Note that, due to symmetry reasons we always find pairs of interference terms with interchanged $n_i$ and $n'_i$, which give identical contributions to the four-index sum. In figures where individual terms are shown, they thus overlay each other and are only depicted by a single line. If the sum over all terms is performed these lines need to be counted twice.
Besides, any term that would be allowed by energy conservation (enforced by the $\delta$-function in \equref{partcontrib}) but %, due to being of higher orders in $\zeta$, 
is several orders of magnitude smaller than the strongest terms will not be depicted.

%Throughout the results we apply \emph{atomic units} by setting $\hbar = m = e = 1$. Particularly the atomic unit of a frequency is given by $\text{au} = \nicefrac{v_0}{a_0} = \sci{4.13}{16}\unit{s}^{-1}$, with the Bohr radius $a_0$, the velocity of an electron on the first Bohr orbit $v_0 = \alpha c$ and the fine-structure constant $\alpha$.
% = \sci{0.528}{-8} \unit{cm}
% = \sci{2.18}{8}~\nicefrac{\text{cm}}{\text{s}}

Finally, we would like to note that for all results in \secref{results} a proton beam target is assumed by setting $Z = 1$.
As long as our first-order treatment of the nuclear field does not need Coulomb corrections, pair-creation rates for higher nuclear charges can be inferred from the proton target results by multiplying with an overall scaling factor of $Z^2$.
The rates in \secref{results} will be given in \emph{atomic units} where $1 \unit{au} = \sci{4.13}{16}\unit{s}^{-1}$.

\subsection{Intensity parameters} \seclab{intensities}

To study the effects of combining two modes in a laser field, we have to ensure both modes have a sizeable influence on the results.
In order to prevent one mode from dominating over the other, which would correspond to an effectively monochromatic laser wave, it is favourable to choose the intensity parameters so that the direct terms for the two modes contribute equally. Consequently the contribution strength of all mixed terms will be maximized.
%
%In order to maximize the contribution strength of the interference terms we choose the intensity parameters so that the direct terms for the two modes contribute equally.
%
Provided that $\xi_i \ll 1$, the $\xi_i$-scaling of the direct term of mode $i$ can be given as
\begin{equation} \equlab{xi-direct}
R_i \sim \xi_i^{2 \tilde{n}_i},
\end{equation}
where the abbreviations $R_1 = R_{[\tilde{n}_1,\tilde{n}_1,  0,  0]}$ and $R_2 = R_{[0  ,0  ,\tilde{n}_2,\tilde{n}_2]}$ are used.
To achieve $R_1 \approx R_2$ one can choose a common parameter $\zeta$ so that the two laser intensity parameters $\xi_i$ (compare \equref{laserwave}) are connected to the other wave's minimal photon number $\tilde{n}_i$ by
\begin{equation}	\equlab{xi-approx}
\xi_1 \approx \zeta^{\tilde{n}_2}
\quad\text{and}\quad
\xi_2 \approx \zeta^{\tilde{n}_1},
\end{equation}
leading to
\begin{equation}	
R_1 \approx R_2 \sim \zeta^{2 \tilde{n}_1 \tilde{n}_2}.
\end{equation}
This definition only takes the general scaling into account and due to differences in the respective proportionality factors a small adjustment has to be made to gain fully equal direct terms in our results. This is achieved by calculating the fully-integrated partial rates of the two direct terms once and inferring the corrected $\xi_i$ from their ratio.

For a general term $[n_1, n'_1, n_2, n'_2]$ a scaling corresponding to \equref{xi-direct} can be given and the individual $\xi_i$ can be replaced by the newly introduced $\zeta$:
\begin{equation}	\equlab{any-rate-xi}
R_{[n_1,n'_1,n_2,n'_2]} \sim \xi_1^{(n_1+n'_1)}  \xi_2^{(n_2+n'_2)} \approx \zeta^{\tilde{n}_2 (n_1+n'_1) + \tilde{n}_1 (n_2+n'_2)}.
\end{equation}
%%
%\begin{align}	
%R_{[n_1,n'_1,n_2,n'_2]} \equlab{any-rate-xi}
%&= \xi_1^{(n_1+n'_1)}  \xi_2^{(n_2+n'_2)} \\
%&\approx \zeta^{\tilde{n}_2 (n_1+n'_1) + \tilde{n}_1 (n_2+n'_2)}.
%\end{align}
%%
%

%\newpage
\section{Results and Discussion} \seclab{results}

%\subsection{Variation of the Total Photon Energy in the Nuclear~Rest~Frame}
%\subsection{Variation of the Relative Phase in the Nuclear~Rest~Frame}
%\subsection{Variation of the Relative Phase in the Laboratory~Frame}

\subsection{Interference and Phase Effects} \seclab{commen}

At first we consider the case of commensurable frequencies.
In \cite{augustin} we used laser pair $(2,4)$ as prime example as it showed the strongest contribution from interference terms. There we compared it explicitly to laser pair $(1,2)$, which has a lower photon number but identical frequency ratio. It is interesting to also study laser pair $(4,8)$ where the frequency ratio is again identical but the total number of photons is higher.

\begin{figure}[t]
\begin{center}

\subfigure[\figlab{const24}
Laser pair $(2,4)$ with $\xi_1 = 10^{-4}$ and $\xi_2 = \sci{1.46}{-2}$
]{%
\includegraphics[scale=0.95]{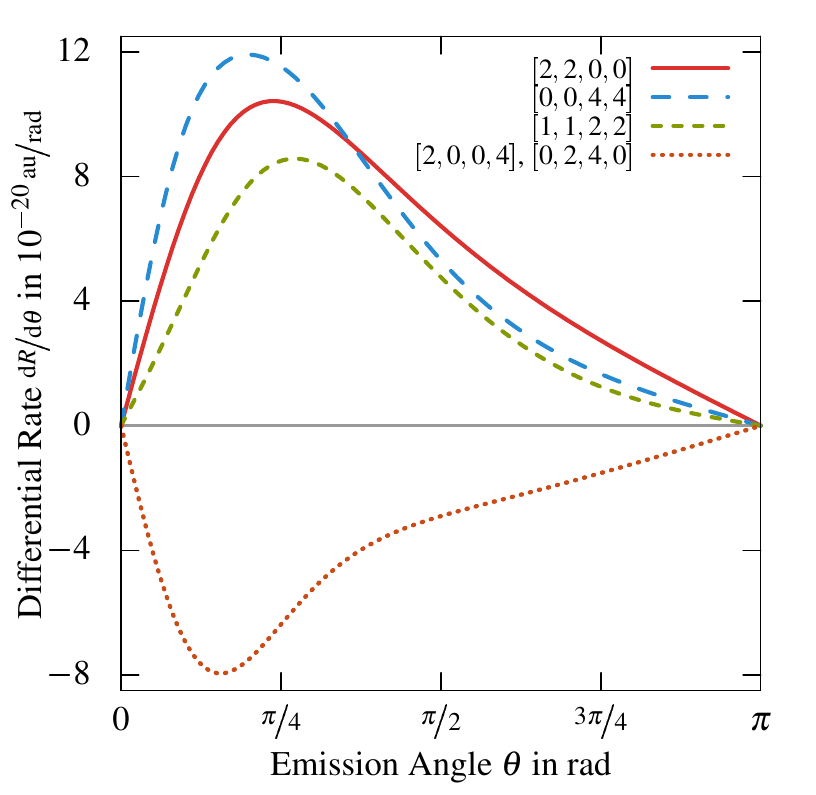}%
}
\subfigure[\figlab{const48}
Laser pair $(4,8)$ with $\xi_1 = 10^{-4}$ and $\xi_2 = \sci{1.31}{-2}$
]{%
\includegraphics[scale=0.95]{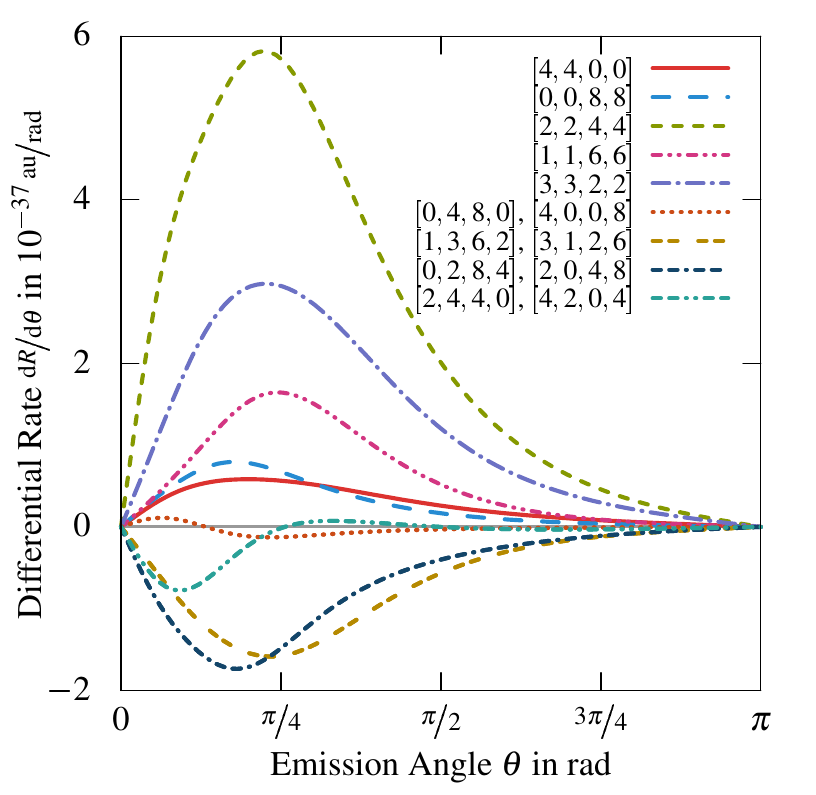}%
}

\caption{\figlab{const2448}(Color online)~
Angular-differential partial rates for laser pairs $(2, 4)$ and $(4, 8)$ with total photon energy $1.15 \unit{MeV}$ in the nuclear rest frame, corresponding to (a) $\omega_1 = 575 \unit{keV}$, $\omega_2 = 287.5 \unit{keV}$ and  (b) $\omega_1 = 287.5 \unit{keV}$, $\omega_2 = 143.75 \unit{keV}$.
The frequencies are chosen such that the total photon energy is reached for $\tilde{n}_1$ photons from the first mode and $\tilde{n}_2$ photons from the second mode.
The emission angle $\theta$ is measured with respect to the laser propagation direction.
The relative phases are set as $\varphi_1 = \varphi_2 = 0$.
}

\end{center}
\end{figure}

In \figref{const2448} a comparison between the laser pairs $(2,4)$ and $(4,8)$ is shown. Due to the higher photon number in the latter example, a larger set of contributing terms needs to be discussed. For $(2,4)$, there is one direct term for each mode ($[2,2,0,0]$ and $[0,0,4,4]$), one symmetrically mixed term $[1,1,2,2]$, using half of the needed energy from each mode, and two interference terms ($[2,0,0,4]$ and $[0,2,4,0]$). All five contributing terms are of identical order in the common parameter $\zeta$ and, consequently, they give similar absolute contributions.

For laser pair $(4,8)$, the corresponding set of terms is again clearly visible: The two direct terms $[4,4,0,0]$ and $[0,0,8,8]$, the symmetrically mixed term $[2,2,4,4]$, and the two interference terms $[4,0,0,8]$ and $[0,4,8,0]$. In contrast to the former case, the symmetrically mixed term $[2,2,4,4]$ is now the strongest contribution. Additionally, we find two more interference terms, $[1, 3, 6, 2]$ and $[3, 1, 2, 6]$, with $n_1 + n'_1 = 4$ and $n_2 + n'_2 = 8$.
Finally, two symmetrically mixed terms with different $n_i + n'_i$ are visible: $[1,1,6,6]$ and $[3,3,2,2]$. Both have corresponding interference terms: $[0, 2, 8, 4]$, $[2, 0, 4, 8]$ and $[2, 4, 4, 0]$, $[4, 2, 0, 4]$, respectively.
It is interesting to note, that all non-negligible contributions are again of the same order in $\zeta$, despite the differing $n_i + n'_i$ in the latter.
%
%A small difference actually manifests in the total intensity available for the respective terms, due to the small adjustment to the individual $\xi_i$ needed to ensure equally contributing direct terms (cf. \equref{any-rate-xi}).
%
Small differences in the contribution strengths remain due to the different prefactors contained in each term (cf. \secref{intensities}).

We find that for laser pair $(4,8)$ the two direct terms are only weak contributions and all symmetrically mixed terms are stronger.
Of the interference terms, the straightforward ones $[4,0,0,8]$ and $[0,4,8,0]$ are weakest, while the more complex ones, with only one or none index equal to zero, are quite strong. This is especially interesting in the comparison of $[1,1,6,6]$, which is the weakest of the three symmetrically mixed terms, and the interference terms with identical $n_i + n'_i$, $[0, 2, 8, 4]$ and $[2, 0, 4, 8]$, which are the strongest of that type.

\begin{figure}[t]
\begin{center}

\includegraphics[scale=0.95]{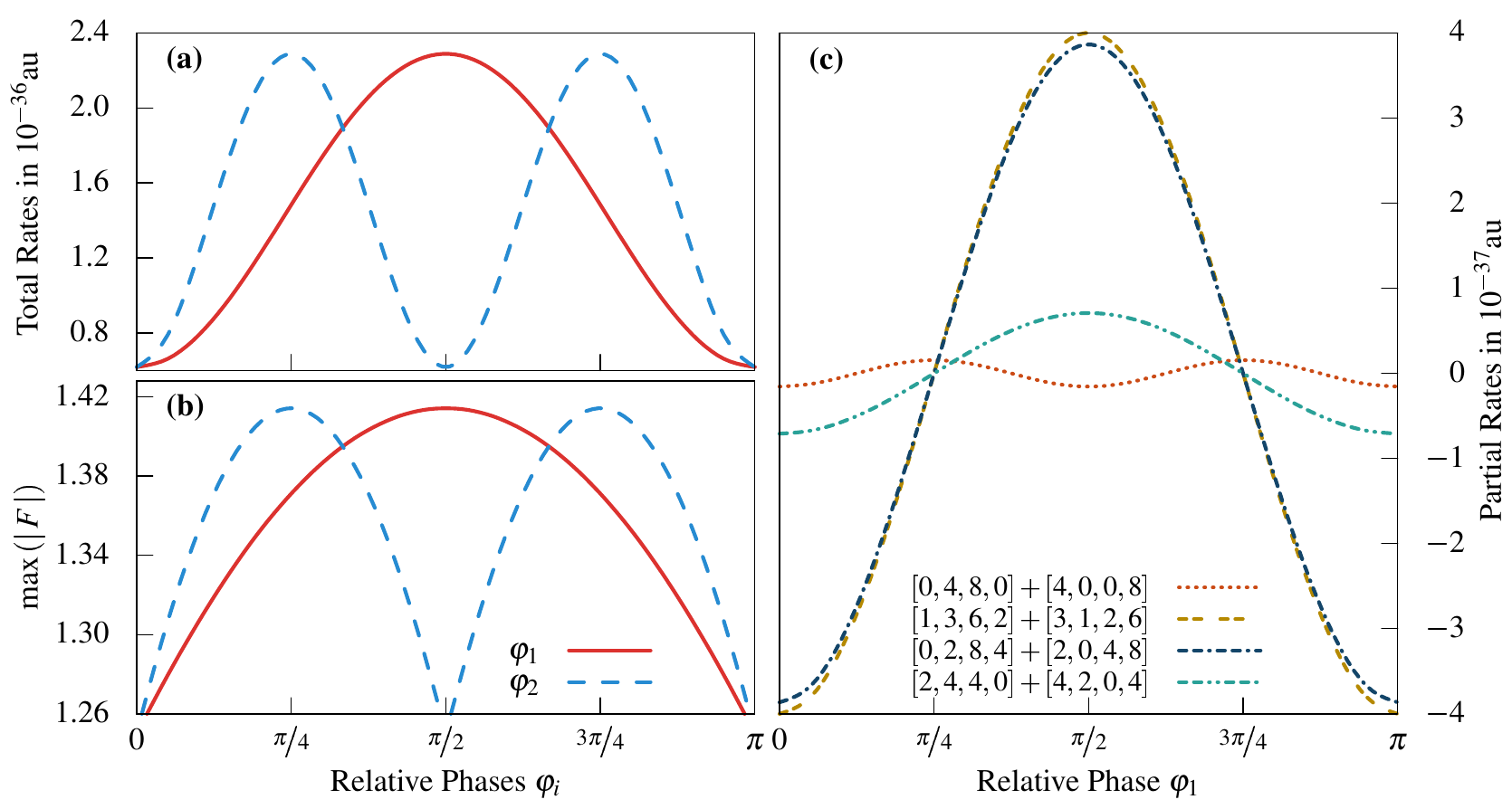}%

\caption{\figlab{phase48}(Color online)~
Variation of the relative phases $\varphi_1$ and $\varphi_2$ for laser pair $(4,8)$ as in \subfigref{const2448}{b} -- 
Comparison of (a) the total pair-production rate and (b) the measure defined via the absolute maximum of the sum of the normalized electric fields from \equref{normelectricfield}. Additionally, in (c) the relevant partial rates of the interference terms are shown for the variation of $\varphi_1$. Note the different periodicity of the terms $[4,0,0,8]$ and $[0,4,8,0]$.
}

\end{center}
\end{figure}

Of all terms shown in \figref{const2448}, only the interference terms are affected by a variation of the relative phase between the two laser modes.
%Varying the phase between the two laser modes will only affect the interference terms.
For laser pair $(2,4)$, their periodicity $\Phi_i$ in the relative phases $\varphi_i$ can be given explicitly as \cite{augustin}
\begin{equation} \equlab{phase-dep}
\Phi_i = \nicefrac{2 \pi}{\tilde{n}_i}.
\end{equation}
The phase dependence can be related to the modulus of the electric field of the combined laser modes.
The measure $\max(\abs{F})$, derived from the electric fields $E_i$ using an appropriate scaling of the mode intensities, is defined via the squared quantity
\begin{equation} \equlab{normelectricfield}
F^2 (ct-z)
= \sum_{i=1}^{2} \frac{c^2}{\vec{a}_i^2 \omega_i^2} E_i^2
= \sum_{i=1}^{2} \sin^2(\eta_i + \varphi_i),
\end{equation}
and exhibits qualitatively the same phase dependence as the total pair-production rate.
%
%A comparison is shown in \figref{phase48} (a) and (b).

%Interestingly, a deeper look into the partial contributions of laser pair $(4,8)$ shows however that not all interference terms share this behaviour (cf. \subfigref{phase48}{c}). Particularly, the terms $[4,0,0,8]$ and $[0,4,8,0]$ have a phase dependence of $\nicefrac{\pi}{2}$ and $\nicefrac{\pi}{4}$ in $\varphi_1$ and $\varphi_2$, respectively.

%Therefore, the periodicity $\Phi_i$ given in \equref{phase-dep} is altered. From the analytical calculations, we find that a more general expression for the periodicity in $\varphi_i$ can be given as $\nicefrac{2 \pi}{\Delta n_i}$ using the difference of the indices of the respective laser mode $i$: $\Delta n_i = \abs{n_i - n'_i}$. The $\Delta n_i$ are indeed identical for the six stronger interference terms and provide the correct phase dependence when replacing $\tilde{n}_i$ in \equref{phase-dep}. Obviously, for interference terms with an index combination similar to $[0,4,8,0]$, i.e., with one vanishing index for each mode, both $\tilde{n}_i$ and $\Delta n_i$ are equal.

The connection between the phase dependences of $\max(\abs{F})$ and the total pair-production rate also holds for laser pair $(4, 8)$, as \subfigsref{phase48}{a} and (b) show. However, \equref{phase-dep} does not apply to this case but requires a generalization. From the analytical calculations, we find that a more general expression for the periodicity in $\varphi_i$ can be given as
\begin{equation} \equlab{phase-dep-gen}
\Phi_i = \nicefrac{2 \pi}{\Delta n_i}
\end{equation}
using the difference of the indices of the respective laser mode $i$: $\Delta n_i = \abs{n_i - n'_i}$. An illustration of \equref{phase-dep-gen} is given in \subfigref{phase48}{c} where the phase dependences of the interference terms of laser pair $(4, 8)$ are shown. Interestingly, not all of them share the same behaviour.
%
%The six stronger interference terms have $\Delta n_1 = 2$ and, thus, exhibit the same periodicity of $\Phi_1 = \pi$ like $\max(\abs{F})$.
%
The six stronger interference terms share $\Delta n_1 = 2$ and accordingly they all exhibit the same periodicity $\Phi_1 = \pi$, which is also the periodicity of $\max(\abs{F})$.
%
%In contrast, the phase dependence of the weakest terms $[4,0,0,8]$ and $[0,4,8,0]$ corresponds to $\Phi_1=\pi/2$, in accordance with $\Delta n_1 = 4$.
%
In contrast, the weakest terms $[4,0,0,8]$ and $[0,4,8,0]$ show a different periodicity of $\Phi_1=\pi/2$, in accordance with $\Delta n_1 = 4$.

Thus we can conclude that, for laser pair $(4, 8)$, those interference terms which exhibit a different phase dependence than $\max(\abs{F})$ are contributing only marginally, while all strong interference terms share the periodicity of this parameter. It is thus retained by the sum of all terms and, eventually, by the total rate.
For laser pair $(2, 4)$ the situation is simpler because all interference terms that give relevant contributions to the total pair-production rate satisfy $\Delta n_i = \tilde n_i$, in agreement with \equref{phase-dep}.

\subsection{Commensurable vs. Incommensurable Frequencies} \seclab{incomm}

In \secref{commen} we have discussed examples of two laser modes with commensurable frequencies $\omega_1$ and $\omega_2$, where $\tilde{n}_1 \omega_1 = \tilde{n}_2 \omega_2$. As, in this case, it is indistinguishable whether $\tilde{n}_1$ or $\tilde{n}_2$ photons were taken from mode one or two, respectively, quantum paths may interfere, giving rise to the effects discussed in said section.

These examples should principally be distinguished from those with incommensurable frequencies, as there interference cannot occur. 
However, due to two arguments, one physical and the other mathematical, this strict distinction may not be an ideal choice (a similar line of argument can be found in \cite{potvliege}):
%
%On the one hand, all physical systems inherently have finite lifetimes, thus measured energies will have a bandwidth, i.e., a laser pulse with a finite pulse length will consist of a range of frequencies, and physical properties should be independent of small variations therein. 
%
On the one hand, laser fields used in a real experimental situation have a finite pulse length and, thus, comprise a continuous range of frequencies; physical properties should possess a smooth dependence under small variations therein.
On the other hand, any irrational number can be approximated by a ratio of integers with arbitrary precision.
Combining these arguments it becomes obvious that commensurability can be understood as always partially fulfilled and thus not allowing the intended distinction.

A physically meaningful replacement is whether the frequency ratio is comprised of two \emph{small} integers or not.
%
%For a small-integer frequency ratio it occurs more frequently that in the superposition of the two laser modes a maximum of one mode coincides with a maximum of the other mode.
%
%Thus the maximally achievable intensity will occur often and repeatedly.
%
%On the other hand, for those ratios with at least one large integer, two coinciding maxima are very rare.
%
For a small-integer frequency ratio the maximally achievable field strength in the superposition of the two laser modes will occur often and repeatedly.
For example, it may happen that the maxima of the mode with the slower oscillation always coincide with a maximum of the other one.
In contrast, for a large-integer frequency ratio, two coinciding maxima are very rare.
Furthermore, the latter case also means that a high number of photons is needed for an interference term, making it less likely.
The rational approximation of an irrational number will certainly consist of two large integers (thus implying minor contributions from interference) and consequently, incommensurable frequencies are treated properly by this distinction.

\begin{figure}[t]
\begin{center}

\includegraphics[scale=0.95]{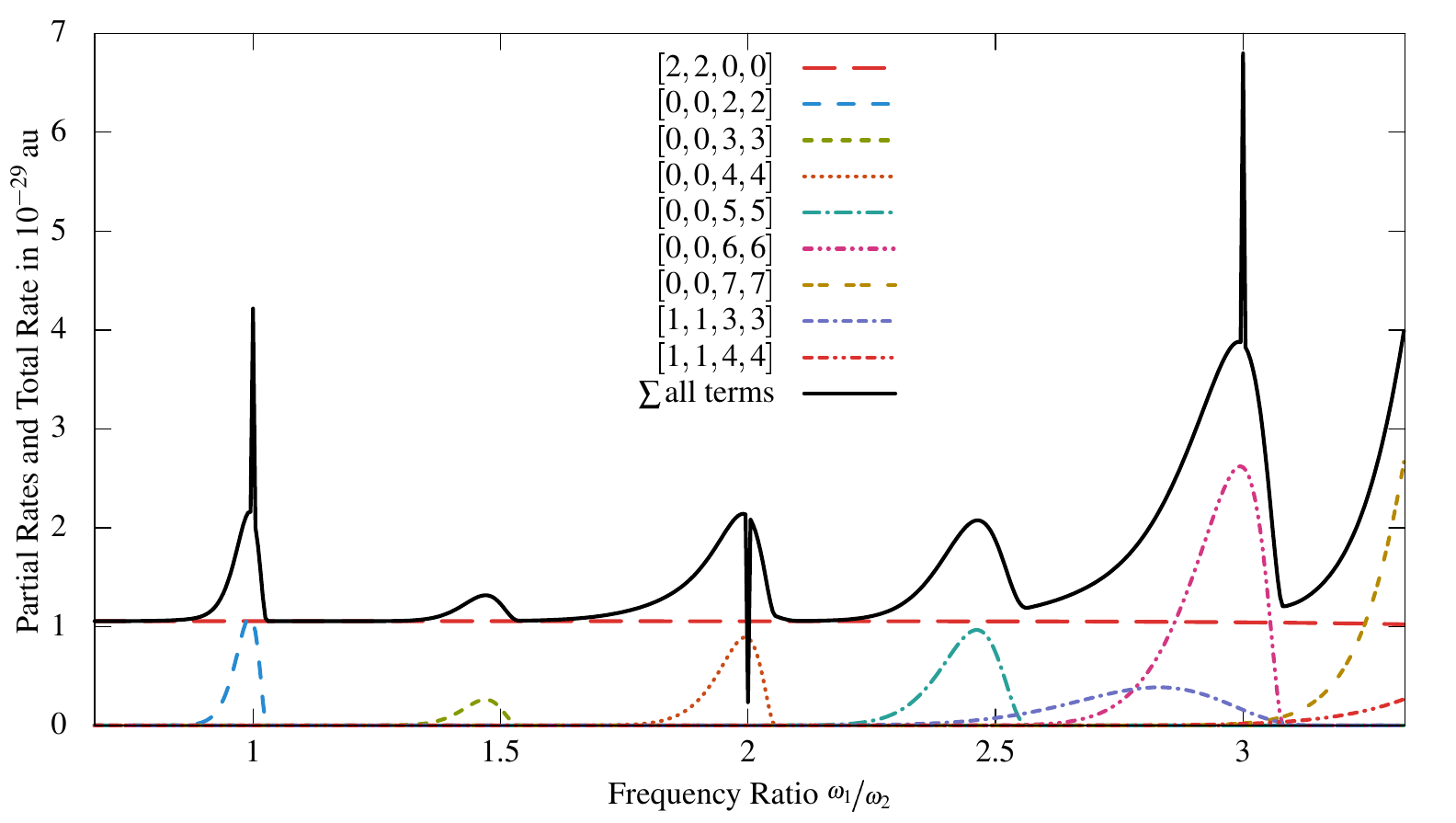}%

\caption{\figlab{varcommen}(Color online)~
Variation of the frequency ratio $\nicefrac{\omega_1}{\omega_2}$ of the two laser modes -- 
Total rate and integrated partial rates for the total photon energy $E_\text{tot} = 1.05 \unit{MeV}$ in the nuclear rest frame
with $\omega_1 = \nicefrac{E_\text{tot}}{2}$ fixed and $\omega_2$ varied from $\nicefrac{E_\text{tot}}{1.8}$ to $\nicefrac{E_\text{tot}}{6.2}$.
%
%The frequencies $\omega_i = \nicefrac{1.15 \unit{MeV}}{\tilde{n}_i}$ are chosen such that the total photon energy is reached for $\tilde{n}_1$ photons from the first mode and, due to $\tilde{n}_2$ not being an integer here, $\ceil{\tilde{n}_2\!}$ photons from the second mode.
%
The relative phases are set as $\varphi_1 = \varphi_2 = 0$.
The intensity parameter $\xi_2 = \sqrt{\nicefrac{\omega_1}{\omega_2}} \zeta^{\nicefrac{\omega_2}{\omega_1}}$ is scaled similar to \equref{xi-approx}, with $\xi_1 = \zeta = 10^{-6}$ and an empirically found prefactor of $\sqrt{\nicefrac{\omega_1}{\omega_2}}$ taking the small adjustment explained in \secref{intensities} into account.
Note that the partial contributions of the interference terms are not shown separately as they are clearly visible as the $\delta$-spikes for the integer frequency ratios.
}

\end{center}
\end{figure}

Applying this idea, we can study a continuous variation of the frequency ratio $\nicefrac{\omega_1}{\omega_2}$, passing by the commensurable laser pairs $(2,4)$ and $(2,6)$ discussed in the earlier study \cite{augustin}, and thus gain further insights into the processes leading to the interferences.
In \figref{varcommen} this variation is plotted by keeping $\omega_1$ fixed and varying $\omega_2$ for the total pair-production rate and several summed-up partial rates.
The spectrum is comprised of the constant direct term of the first mode, $[2,2,0,0]$, and slowly rising and falling peaks for the direct terms of the second mode, $[0,0,n,n]$ for $n = (2,3,4,5,6)$.
Note that, the increase of these direct terms around their respective $\tilde{n}_2$ value can be explained by the applied scaling of the intensity as given in the figure caption.
On top of these peaks, for those commensurable frequencies where both $\tilde{n}_i$ are even integers, constructive ($\tilde{n}_2 = 2$ or $6$) and destructive ($\tilde{n}_2 = 4$) interference terms lead to prominent $\delta$-spikes, which are -- in our treatment -- infinitely narrow but of finite height. Note that the interference terms are not shown separately, but the spike they contribute can be clearly seen in the sum of all terms.

We note that in an actual experimental realisation of the effects studied here, the laser would not be an infinite plane wave but a finite pulse. This would lead to the $\delta$-spikes being smeared out and the resulting spectrum being continuous.
Nevertheless, the strong enhancement or depletion at the even-numbered commensurable frequencies can be expected to remain.

A variation of the relative phases $\varphi_i$ will only affect the spikes, sinusoidally changing their height and sign with the periodicity as discussed in \secref{commen}. This has been examined extensively in the aforementioned earlier study, particularly for the laser pairs $(2, 4)$ and $(2, 6)$. In the next section, we would like to concentrate on the special case $(2,2)$ instead, representing a single linearly polarized or a circularly polarized laser wave for the difference of the relative phases $\abs{\varphi_1 - \varphi_2}$ set to $0$ or $\nicefrac{\pi}{2}$, respectively. Here, a variation of the phase corresponds to a variation of the ellipticity of the combined laser wave.

\subsection{Monochromatic Laser Wave of Elliptical Polarization} \seclab{ellip}

A vector potential of the form given in \equref{laserwave} can also describe a monochromatic laser wave if we set $\omega_1 = \omega_2$.
Particularly, the setup denoted as laser pair $(2,2)$ in the \secref{incomm} corresponds to a single linearly polarized laser wave with the polarization axis lying diagonally in the $x$-$y$-plane for vanishing relative phases $\varphi_1 = \varphi_2 = 0$, and to a circularly polarized laser wave for one $\varphi_i$ set to $\nicefrac{\pi}{2}$ instead. By a continuous variation between these two extreme conditions a laser wave of arbitrary ellipticity can be studied.
The variation may be carried out in two different ways, both starting from a circularly polarized wave:
On the one hand one can fix the maximum amplitude of the laser field and decrease one intensity half-axis until it disappears, while keeping the other half-axis constant. 
On the other hand one can fix the total intensity supplied to the system, which means decreasing one half-axis while increasing the other, so that $\xi_1^2 + \xi_2^2 = \text{const}$.%
\footnote{It can be shown that the latter pathway corresponds directly to the variation of the relative phase mentioned at the end of \secref{incomm} for fixed intensity parameters, which is interesting as $\varphi_i$ and $\xi_i$ are independent parameters. For a variation of $\varphi_i$ the ellipticity is then given as $\varepsilon = \abs{\cos{\varphi_i}}$.}
In both cases the ellipticity will be measured using the parameter
\begin{equation}	
\varepsilon = \frac{\abs{\xi_1^2 - \xi_2^2}}{\xi_1^2 + \xi_2^2} = 
\begin{cases}
0 &\text{for circular polarization} \\
0 < \varepsilon < 1 &\text{for elliptical polarization} \\
1 &\text{for linear polarization.}
\end{cases}
\end{equation}
%
%which equals 1 for a linear polarization and 0 for a circular polarization, 
%can be adjusted to the case of a fixed total intensity:
%%
%\begin{equation}	
%\varepsilon = \frac{2 \xi_1^2}{I_\text{total}} - 1 = 1 - \frac{2 \xi_2^2}{I_\text{total}}.
%\end{equation}
%%

\begin{figure}[t]
\begin{center}

\includegraphics[scale=0.95]{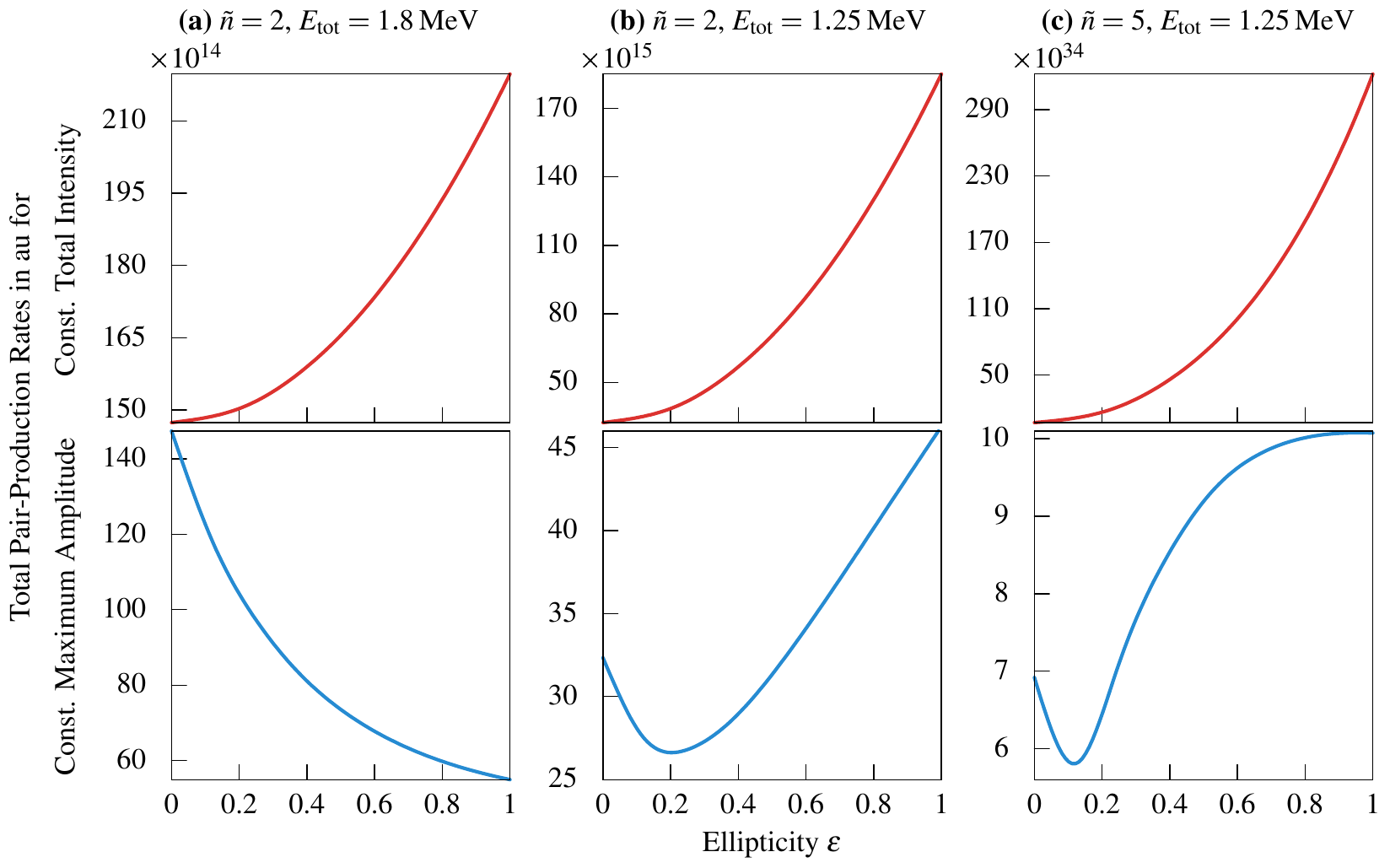}%

\caption{\figlab{ellipt}(Color online)~
Variation of the ellipticity $\varepsilon$ --
Total rates in the nuclear rest frame for laser pair $(2, 2)$ with total photon energies (a) $1.8 \unit{MeV}$ and (b) $1.25 \unit{MeV}$, and for (c) laser pair $(5, 5)$ with total photon energy $1.25 \unit{MeV}$.
Comparisons are shown between (top panels) fixed total intensity using
%
%$\sqrt{\xi_1^2 + \xi_2^2} = \sqrt{2} \left( \sci{7.5}{-4} \right)$,
%
%alt:
%
${\xi_1^2 + \xi_2^2} 
%= {2} \left( \sci{7.5}{-4} \right)^2$,
= \sci{2}{-6}$,
and (bottom panels) fixed maximum laser wave amplitude using
%
%$\xi_1 = \sci{7.5}{-4}$
$\xi_1^2 = 10^{-6}$
and
%
%$\xi_2 = 0, \dots, \xi_1$.
%
%alt:
%
$0\le\xi_2\le\xi_1$.
For the two-photon cases, where it is applicable, the fully analytical graphs obtained using the formulas derived in \cite{milstein} coincide perfectly with our results.
}

\end{center}
\end{figure}

In \subfigref{ellipt}{a} the two pathways are shown in comparison for a total photon energy of $1.8 \unit{MeV}$ and $\tilde{n} = 2$ photons absorbed from the wave.
For the first case of a constant maximum laser wave amplitude (bottom panel), we see a decreasing total pair-production rate, when going from circular to linear polarization. This can be intuitively explained by the decrease of the supplied total laser intensity as we gradually turn off the second laser mode.
For the second case of constant total intensity (top panel), we see an increase from left to right instead. Here the peak amplitude of the combined laser waves allows an explanation. In the circularly polarized case the field amplitude is constant, while only the field vector direction changes:
%%
%\begin{equation}
%%
%\vec{A}_\text{circ}
%= \vec{a}_1 \sin\eta + \vec{a}_2 \sin(\eta + \nicefrac{\pi}{2})
%= a \left( \vec{e}_1 \sin\eta + \vec{e}_2 \cos\eta \right)
%,\qquad
%%
%\abs{\vec{A}_\text{circ}}
%= a
%.
%%
%\end{equation}
%%
%In the linearly polarized case the field amplitude is subject to the sinusoidal variation of the wave, but the peak amplitude is increased by a factor of $\sqrt{2}$ due to the superposition of the two waves:
%%
%\begin{equation}
%%
%\vec{A}_\text{lin}
%= \vec{a}_1 \sin\eta + \vec{a}_2 \sin\eta
%= a \left( \vec{e}_1  + \vec{e}_2 \right) \sin\eta
%,\qquad
%%
%\abs{\vec{A}_\text{lin}}
%= a \sqrt{2} \sin\eta
%.
%%
%\end{equation}
%%
%
%
%
\begin{alignat}{3}
\equlab{amp-circ}
\vec{A}_\text{circ}
&= \vec{a}_1 \sin\eta + \vec{a}_2 \sin(\eta + \nicefrac{\pi}{2})
&&= a \left( \vec{e}_1 \sin\eta + \vec{e}_2 \cos\eta \right)
,\qquad
&\abs{\vec{A}_\text{circ}}
&= a
.\\
\intertext{In the linearly polarized case the field amplitude is subject to the sinusoidal variation of the wave, but the peak amplitude is increased by a factor of $\sqrt{2}$ due to the superposition of the two waves:}
\equlab{amp-lin}
\vec{A}_\text{lin}
&= \vec{a}_1 \sin\eta + \vec{a}_2 \sin\eta
&&= a \left( \vec{e}_1  + \vec{e}_2 \right) \sin\eta
,
&\abs{\vec{A}_\text{lin}}
&= a \sqrt{2} \sin\eta
.
\end{alignat}

The explanation given for the ellipticity dependence for a fixed total field intensity also applies to lower total photon energies and higher photon orders. Corresponding examples are shown in the top panels of \subfigsref{ellipt}{b} and (c) for a total photon energy of $1.25 \unit{MeV}$ and $\tilde{n}=2$ or $5$ absorbed photons, respectively. However, remarkably, the argument provided before for the ellipticity dependence for a fixed maximum field amplitude is not applicable to these parameters. As the bottom panels of \subfigsref{ellipt}{b} and (c) illustrate, the rate exhibits a non-monotonous dependence on the ellipticity here. After passing through a minimum at about $\varepsilon \approx 0.2$ and $0.12$, respectively, the rate starts growing again and reaches its maximum value for a linearly polarized wave. The reason for this striking difference seems to be related to the excess energy above the pair creation threshold $\Delta E = \tilde{n} \omega - 2 m c^2$, which is much smaller here than in \subfigref{ellipt}{a}. We note that a non-monotonous dependence on the applied field ellipticity has also been obtained recently for the rate of pair production by the nonlinear Breit--Wheeler process \cite{heliangyong}.

Bethe--Heitler pair creation by an elliptically polarized, monochromatic laser wave has also been studied in \cite{milstein} using a polarization-operator approach. In particular, analytical expressions for the total pair-production rate by two-photon absorption were obtained in Eqs.\,(15) and (26) therein.
We stress that the ellipticity dependences following from these expressions coincide perfectly with our numerically calculated results for the two-photon case in \subfigsref{ellipt}{a} and (b) (top and bottom panels). Moreover, our present approach allows us to extend these results straightforwardly to higher photon orders. The five-photon case shown in \subfigref{ellipt}{c} serves as an example. It exhibits features qualitatively similar to the two-photon case.

%\newpage
\section{Summary and Conclusion} \seclab{concl}
We have studied electron-positron pair creation on a nucleus by multiphoton absorption from an intense two-mode laser field. If the laser field contains two different frequencies of commensurable ratio we found quantum interference effects to be visible in the angular distribution of the produced particles as well as in the total production rate. The latter can be explained within an intuitive picture based on the phase dependence of the maximum electric field amplitude. Additionally, the case of two incommensurable frequencies was examined by continuously varying their ratio. Finally, for identical frequencies of both field modes, the dependence of the pair production rate on the ellipticity of the laser wave was studied. While this dependence shows the expected behaviour when the field intensity is held constant, interesting features were found when the maximum field amplitude is kept fixed instead.

Our results could be tested experimentally, for instance, by combining a relativistic proton beam of Lorentz factor $\gamma \sim 50$ with a counterpropagating intense X-ray laser beam comprising two field modes \cite{lutman}.

\ack
Funding for this project by the German Research Foundation (DFG) under Grant No. MU 3149/1-1 is gratefully acknowledged.
S.\,A. also wishes to thank the Heidelberg Graduate School of Fundamental Physics (HGSFP) for the generous travel support.

%\newpage
\section*{References}
%\nocite{*}
\bibliography{b1nourl}

%\the\textwidth
\end{document}